\documentclass[runningheads]{llncs}

\usepackage[T1]{fontenc}
\usepackage{graphicx,verbatim}
\usepackage{amsmath,amssymb}
\usepackage{array}
\usepackage{booktabs}
\usepackage{xcolor}
\usepackage{multirow}
\usepackage{bbding}
\begin{document}

\title{CaseWeaver: A Multi-Agent Framework for Multimodal Virtual Clinical Case Generation}
\titlerunning{CaseWeaver: Multimodal Virtual Clinical Case Generation}

\author{Jierui Qu\inst{1} \and Jiachuan Peng\inst{1,2} \and
Lin Li\inst{3} \and Kyle Lam\inst{4} \and Jianing Qiu\inst{1}\inst{(}\Envelope\inst{)}}
\authorrunning{J. Qu et al.}
\institute{Biological and Life Sciences Division, MBZUAI, UAE\\
\email{Jianing.Qiu@mbzuai.ac.ae}
\and Department of Engineering Science, University of Oxford, UK
\and OATML, Department of Computer Science, University of Oxford, UK
\and Department of Surgery and Cancer, Imperial College London, UK}

\maketitle

\begin{abstract}
Clinical diagnosis relies on consistent multimodal data collected from the same patient throughout the disease course, yet such data are difficult to acquire at scale because of collection costs, missing modalities, fragmented systems, and longitudinal follow-ups. Existing synthetic-data approaches largely focus on individual modalities or vision-language dual modalities at report-level generation. Little work has been done to construct synthetic data with consistent patient backgrounds, coherent disease trajectories, and interrelated modality-specific evidence at a complete clinical case level. We introduce CaseWeaver, a multi-agent framework built around a timeline-anchored Latent Clinical Case Graph (LCCG). The LCCG organizes patient context, latent disease states, clinical events, and expected observations in a shared patient-level representation. Modality-agents use scoped observation subgraphs and clinical protocols to generate evidence including clinical records, laboratory results, physiological signals, and medical images. We evaluate clinical inferability using a calibrated AgentClinic protocol and case diversity using Virtual Case Diversity (VCD) score. CaseWeaver outperformed general-model and agentic-workflow baselines on both metrics, producing more diverse and coherent multimodal virtual clinical cases.

\end{abstract}

\section{Introduction}

Longitudinal multimodal data across a patient’s baseline status and evolving disease course are crucial for supporting patient-centred clinical care \cite{soenksen2022integrated}. However, data collected in practice are shaped by each patient’s pre-existing medical history and specific care pathway and are often fragmented across healthcare systems, resulting in substantial modality missingness in real-world multimodal cohorts \cite{mota2024mmist}. High-quality multimodal data are also difficult to obtain for medical AI training and clinical education. Deep-phenotyping cohort initiatives such as the Human Phenotype Project demonstrate that collecting such data at scale requires substantial effort and resources \cite{reicher2025deep}, while sharing patient data raises privacy concerns \cite{marino2025medical}. Moreover, existing medical datasets may underrepresent particular populations, risking the propagation of demographic and geographic biases into medical AI systems \cite{alderman2025tackling}. Virtual cases with coherent multimodal evidence have the potential to complement real-world cohorts and provide an additional resource for medical AI training and clinical education \cite{brugge2024large}.

Large language models (LLMs) are increasingly used for synthetic data generation \cite{zhong2024memorybank}. In medicine, this trend has extended across multiple data modalities. Studies have used LLMs to generate medical text, clinical records, and structured data \cite{xu2024knowledge,kweon2024publicly}. Generative models have also been used to synthesize physiological time series, including electrocardiograms, electroencephalograms, and continuous glucose monitoring data \cite{neifar2024diffecg,lutsker2026foundation}. Conditional diffusion models generate medical and pathology images from text, disease states, or acquisition conditions \cite{qiu2024development,wang2025selfimproving,han2026beyond}, while general-purpose image models are increasingly being explored for medical illustration, image editing, and disease-specific synthesis \cite{peng2025recognizing}. Despite these advances, most methods remain focused on a single data type or narrow generation objectives. Even when preserving within-modality temporal structure, they lack a unified patient-level representation capable of organizing heterogeneous clinical evidence into a coherent patient world.

To address these limitations, researchers have begun to explore cross-modal clinical data generation. One line of work conditions the generation of one modality on existing modalities from specific real-world datasets. For example, MedM2G supports cross-modal generation between text and images and across CT, MRI, and X-ray \cite{zhan2024medm2g}. However, such methods remain constrained by the disease, anatomical, modality, and acquisition distributions of their source datasets, limiting their coverage of diverse patient contexts \cite{draghi2021bayesboost,ktena2024generative}. Another line of work focuses on patient generation grounded in clinical logic. For example, Patient-Zero generates patient agents with high clinical quality and interaction fidelity \cite{lai2026patientzero}. However, its multimodal representations remain centered on structured reports and natural-language descriptions, without modality-specific evidence for the same patient, such as physiological waveforms, medical images, or pathology images.

We introduce CaseWeaver, a multi-agent framework built around a timeline-anchored Latent Clinical Case Graph (LCCG) as a shared patient-level representation. It generates virtual clinical cases with consistent patient backgrounds, coherent trajectories, and interrelated multimodal evidence. Grounded in clinical guidelines, a case-planning agent encodes a personalized patient background, disease-course timeline, and multimodal examination plan into the LCCG. Modality agents then receive scoped observation subgraphs and use clinical protocols to generate interrelated clinical records, laboratory results, physiological signals, medical images, and pathology images. By linking these outputs to the same patient state, the LCCG preserves clinical coherence and supports diverse disease phenotypes, backgrounds, and care pathways. We evaluate case diversity and clinical inferability against direct LLM and agent-based generation baselines.

\section{Method}

\begin{figure}[t]
\centering
\includegraphics[width=\textwidth]{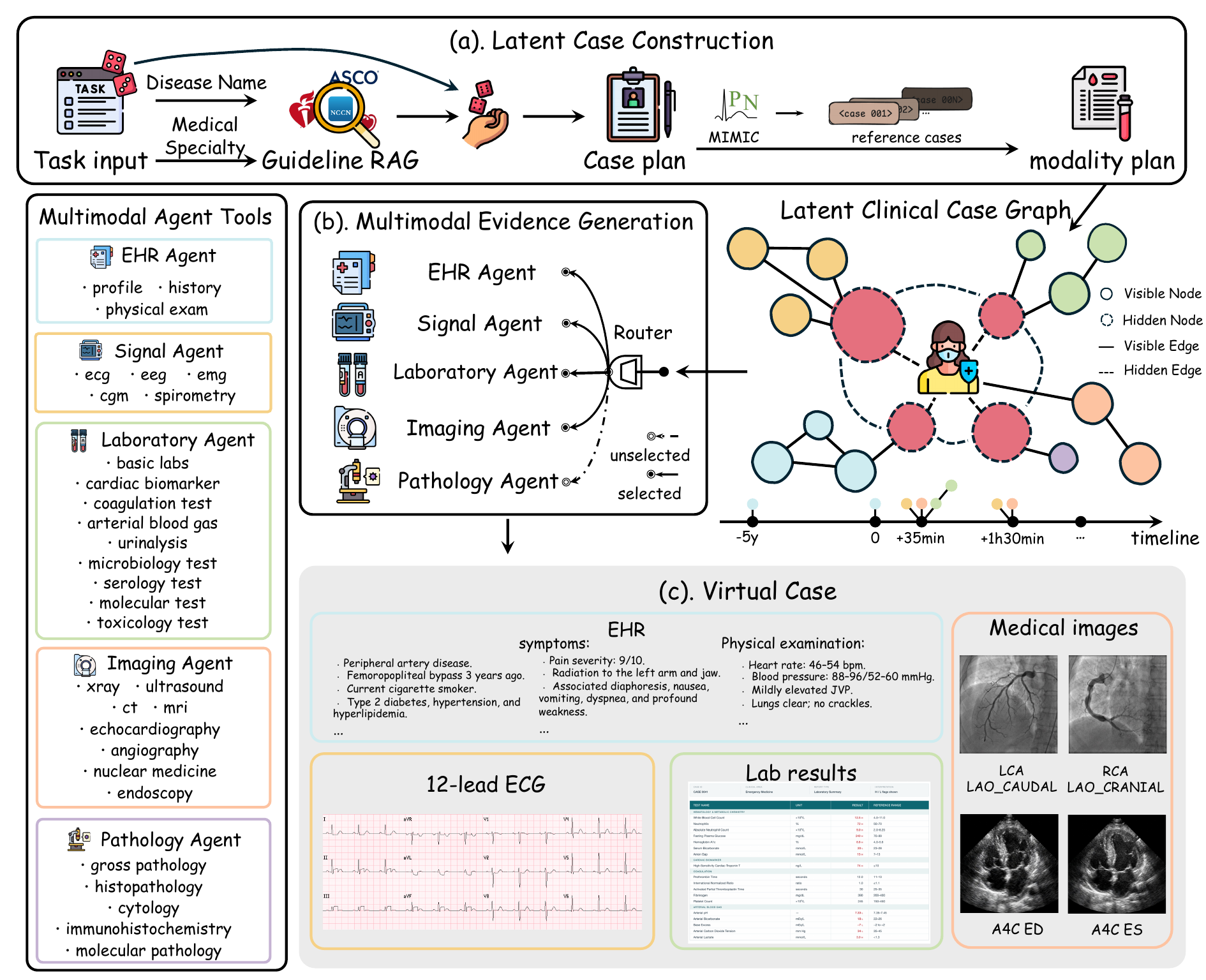}
\caption{Overview of CaseWeaver. (a) Construction of a target-disease-specific LCCG. (b) Generation of multimodal evidence by modality-specific agents. (c) Example of a multimodal virtual clinical case.}
\label{fig:caseweaver-overview}
\end{figure}

CaseWeaver is a multi-agent framework for virtual clinical case generation built around a timeline-anchored LCCG, as illustrated in Fig.~\ref{fig:caseweaver-overview}. Given a task specification comprising a target disease and medical specialty, the framework integrates retrieved clinical knowledge with case and modality plans into the LCCG. It then routes modality-specific observation subgraphs to protocol-constrained agents for evidence generation. The following sections describe latent case construction, multimodal evidence rendering, and evaluation using Virtual Case Diversity and a calibrated AgentClinic-based clinical inferability protocol.

\subsection{Latent Clinical Case Graph Construction}

Given a task specification, CaseWeaver retrieves disease knowledge from authoritative clinical guidelines and summarizes it as a reusable disease program comprising disease subtypes, pathophysiology, clinical course, and recommended modalities. Conditioned on the program and a random seed, the case-planning agent defines the disease presentation, acuity, severity, demographics, risk factors, comorbidities, and relevant clinical history. The seed varies these attributes within guideline constraints to produce distinct configurations for the same disease. Based on the planned demographic and risk-factor profile, the system retrieves a contextually similar, de-identified case from MIMIC-IV~\cite{johnson2023mimiciv} to inform realistic examination planning, while disease mechanisms remain governed by guideline-derived knowledge. Finally, it combines the disease program, case plan, and reference case to select evidence modalities and tools, yielding a case-specific modality plan.

CaseWeaver encodes these outputs in a timeline-anchored LCCG, defined as:

\[
\mathcal{G} = (\mathcal{T}, \mathcal{V}, \mathcal{E}), \qquad
\tau: \mathcal{V} \rightarrow \mathcal{T},
\]

Here, $\mathcal{T}$ denotes clinical time points ordered relative to disease or symptom onset ($t=0$); preceding and subsequent events receive negative and positive times, spanning baseline to outcome. $\mathcal{V}$ contains latent disease and pathophysiology nodes and modality-observable findings, with observable nodes assigned to a modality and tool. Latent nodes are hidden from evidence-generation agents, whereas observable nodes exclude the target diagnosis and equivalent answer cues. $\mathcal{E}$ encodes causal, temporal, evidential (support and counterevidence), co-occurrence, compositional, and cross-modal consistency relations, while $\tau$ anchors each node to $\mathcal{T}$. Every observable finding must be traceable to an underlying disease or pathophysiology node, and cross-modal edges connect findings arising from the same latent patient state. The resulting LCCG ($\mathcal{G}$) jointly represents patient context, disease evolution, modality planning, and expected clinical findings, and serves as input to protocol-driven evidence rendering.

\subsection{Protocol-Driven Multimodal Evidence Rendering}

The LCCG represents the complete latent patient world, whereas each modality-agent should access only information within its scope. For each target modality, the router forms an observation subgraph by selecting agent-visible nodes assigned to that modality and retaining relations whose endpoints are both selected. The subgraph inherits temporal anchors and carries the demographic context and tool assignments required for generation, while excluding latent disease and pathophysiology nodes and observations from other modalities. It is then dispatched to the corresponding agent, which invokes only the tools specified in the modality plan. Additionally, the router organizes observation requirements according to the examination tool assigned to each node, and the five modality-agents can process their respective observation subgraphs in parallel. This routing preserves shared constraints from the patient state and disease-course timeline while enforcing modality-specific information boundaries.

Within each agent, dedicated tools convert natural-language observation requirements into case-specific modality representations through a protocol-driven update (Fig.~\ref{fig:protocol-rendering}). For an observation requirement $x$, the update is:

\begin{figure}[t]
\centering
\includegraphics[width=\textwidth]{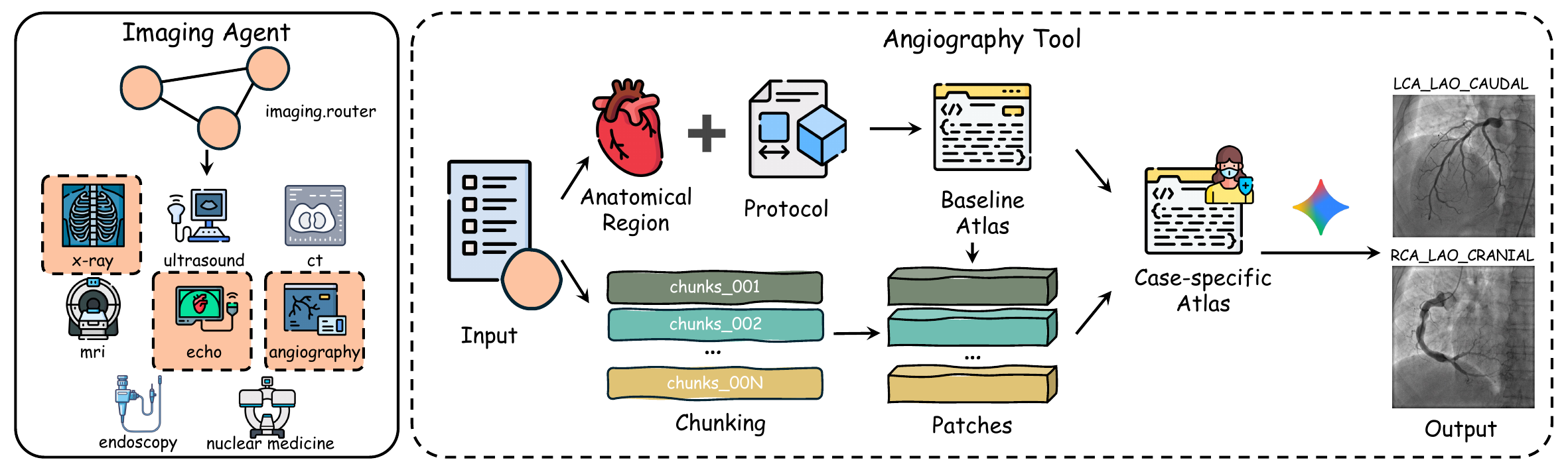}
\caption{Protocol-driven multimodal evidence rendering. The Imaging Agent--Angiography Tool workflow is shown as an example. The tool selects the target coronary anatomy according to task requirements and applies the \texttt{LCA\_LAO\_CAUDAL} (LAO $40^{\circ}$, CAUDAL $30^{\circ}$) and \texttt{RCA\_LAO\_CRANIAL} (LAO $40^{\circ}$, CRANIAL $20^{\circ}$) acquisition protocols.}
\label{fig:protocol-rendering}
\end{figure}

\[
\pi = f_{\mathrm{sel}}(x), \qquad
B_x^{*}
=
B_{\pi}
\oplus
P\left(f_{\mathrm{chunk}}(x),B_{\pi}\right)
\]

Here, $f_{\mathrm{sel}}$ selects a clinical protocol $\pi$, and $B_{\pi}$ denotes its protocol-specific baseline representation. $f_{\mathrm{chunk}}$ decomposes $x$ into clinical findings, $P$ maps them to constrained patches, and $\oplus$ applies the patches to $B_{\pi}$ to produce $B_x^{*}$ which is the case-specific modality representation. A modality renderer then converts $B_x^{*}$ into final evidence. Although tools use different baselines, patch structures, and renderers, this shared update confines case-specific modifications to the selected protocol's representation space, constraining evidence structure, content scope, and modality-specific properties.

\subsection{Evaluation Methodology}

\subsubsection{AgentClinic-Based Clinical Inferability Evaluation}

Generated virtual cases are formatted as diagnostic tasks compatible with AgentClinic's interactive environment, in which a Doctor agent acquires information from Patient and Measurement agents and a Moderator adjudicates the final diagnosis \cite{schmidgall2026agentclinic}. Each case $i$, containing $M_i$ evidence items, is evaluated over $K_i$ independent runs. Its diagnostic success rate $S_i$ and information coverage $C_i$ are calculated as:

\[
S_i
=
\frac{1}{K_i}
\sum_{r=1}^{K_i}
\mathbf{1}\!\left(\hat{y}_{ir}=y_i\right),
\qquad
C_i
=
\frac{1}{K_i}
\sum_{r=1}^{K_i}
\left(
\frac{1}{M_i}
\sum_{j=1}^{M_i}
z_{irj}
\right),
\]

where $r$ indexes independent evaluation runs for case $i$, $y_i$ and $\hat{y}_{ir}$ are the target and predicted diagnoses, respectively, and $z_{irj}=1$ if evidence item $j$ is acquired before the final diagnosis and $0$ otherwise.

To jointly quantify diagnostic closure and information access, we define the Clinical Inferability Index (CII) as the harmonic mean of $S_i$ and $C_i$, and use $G_i$ to quantify their imbalance:

\[
\mathrm{CII}_i
=
\frac{2S_iC_i}{S_i+C_i},
\qquad
G_i=S_i-C_i,
\]

with $\mathrm{CII}_i=0$ when $S_i+C_i=0$. Decision criteria are calibrated using reference cases compiled from real clinical data. The lower bounds for CII and information coverage are set to the fifth percentiles of their empirical distributions in the calibration set, whereas the reference range for $G$ is defined as its central 95\% empirical interval. The fixed criteria are then assessed on the validation set. A case is classified as clinically inferable when its CII and information coverage meet their respective lower bounds and its $G$ falls within the reference range. 

\subsubsection{Virtual Case Diversity}

To quantify the diversity of cases generated for the same task, we define Virtual Case Diversity (VCD). Each case is represented by four feature groups: core phenotype, patient background, clinical course, and evidence plan. Categorical, numerical, set-valued, and textual fields are compared using matching distance, normalized distance, Jaccard distance, and TF--IDF cosine distance, respectively \cite{coombes2021simulation}. The distance between cases $i$ and $j$ is defined as:

\[
d_{ij}
=
\sum_{k=1}^{K} w_k d_k(i,j),
\qquad
d_{ij}\in[0,1]
\]

where $d_k(i,j)$ denotes the normalized distance between cases $i$ and $j$ for the $k$th feature, and $w_k$ is the corresponding feature weight. For a set containing $n$ cases, VCD combines overall pairwise dispersion with nearest-neighbor uniqueness:

\[
\mathrm{VCD}
=
\lambda
\frac{2}{n(n-1)}
\sum_{i<j} d_{ij}
+
(1-\lambda)
\frac{1}{n}
\sum_{i=1}^{n}
\min_{j\ne i} d_{ij},
\qquad
\lambda\in[0,1].
\]

The nearest-neighbor term penalizes duplicate or near-duplicate cases. This metric does not assess the medical correctness of the generated cases.

\section{Evaluation}

During evaluation, all CaseWeaver agents used gpt-5.6-terra. We compared CaseWeaver with four baselines spanning two generation paradigms and two backbone models (gpt-5.6-terra and gemini-3.5-flash). The General Model baselines generated a case plan in a single-turn session, then generated evidence for the $N_{\mathrm{mod}}$ selected modalities through $N_{\mathrm{mod}}$ turns in a separate session. The Agentic Workflow baselines used the OpenAI Agents SDK or Google ADK, with one Planner Agent and five modality agents. For all methods, image evidence was rendered from case-specific prompts using gemini-3-pro-image.

\subsection{Generated Virtual Case}

Fig.~\ref{fig:caseweaver-overview}(c) illustrates a myocardial infarction case generated by CaseWeaver. The generated case was a 68-year-old woman with multiple atherosclerotic risk factors who presented with acute chest pain, autonomic symptoms, and hemodynamic compromise. Electrocardiography showed ST-segment elevation in leads II, III, and aVF, with reciprocal ST depression in leads I and aVL, accompanied by elevated high-sensitivity cardiac troponin T. Echocardiography demonstrated inferior-wall and right-ventricular wall-motion abnormalities, while coronary angiography 90 minutes after symptom onset revealed complete proximal right coronary artery occlusion. These temporally aligned multimodal findings formed a coherent clinical phenotype of inferior ST-segment elevation myocardial infarction with right ventricular involvement.

\subsection{Clinical Inferability Evaluation}

\begin{figure}[t]
\centering
\includegraphics[width=\textwidth]{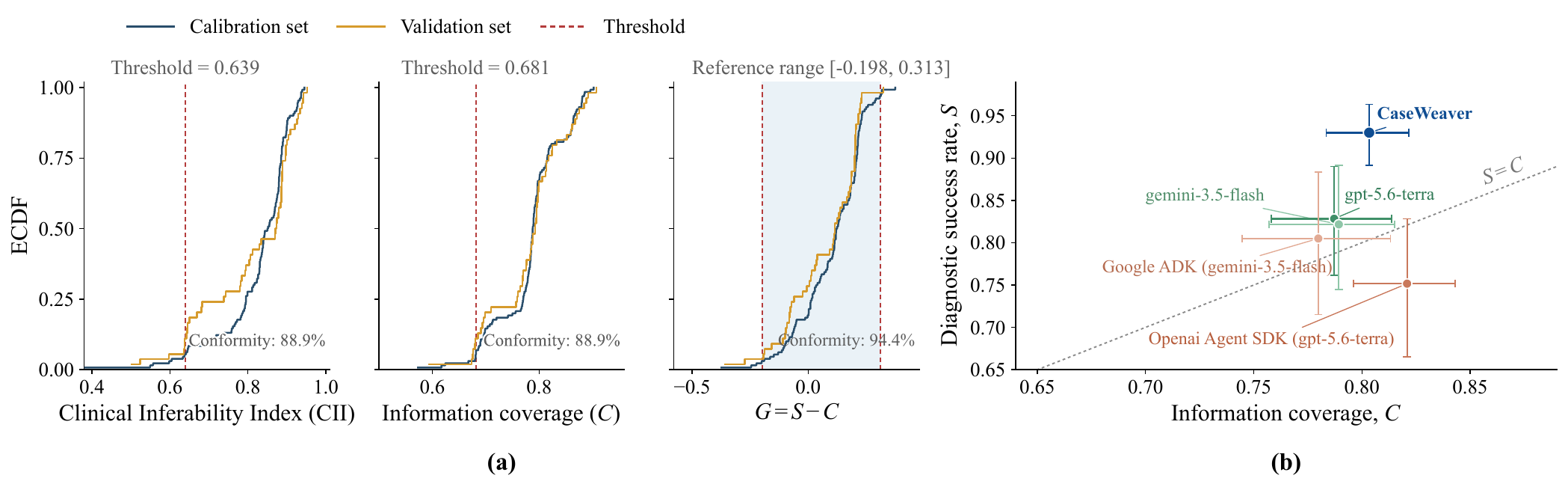}
\caption{(a) Empirical cumulative distributions of CII, information coverage $C$, and the diagnostic success--coverage gap $G$ in the calibration and validation sets; red dashed lines denote the calibration thresholds or reference interval. (b) Joint distribution of the diagnostic success rate $S$ and information coverage $C$.}
\label{fig:clinical-inferability}
\end{figure}

We calibrated the AgentClinic evaluation protocol using screened AgentClinic MedQA reference cases \cite{schmidgall2026agentclinic,jin2021disease}. The calibration and validation sets contained 130 and 54 cases, respectively. The lower bounds for CII and information coverage were 0.639 and 0.681, while the reference interval for $G$ was $[-0.198, 0.313]$ (Fig.~\ref{fig:clinical-inferability}a). In the validation set, 87.0\% of cases conformed to the combined criterion, supporting the stability of the fixed criteria on real reference cases.

For method comparison, we constructed a fixed list of 50 diseases from common diagnoses in AgentClinic MedQA and required CaseWeaver and the four baselines to generate virtual cases for the same disease list. All cases underwent repeated diagnostic evaluation under an identical AgentClinic configuration. CaseWeaver achieved the highest diagnostic success rate ($S=0.930$) and CII (0.860) among all evaluated methods (Table~\ref{tab:clinical-vcd-results}).

\begin{table}[!t]
\centering
\caption{Comparison of clinical inferability and VCD across virtual case generation methods.}
\label{tab:clinical-vcd-results}
\begingroup
\fontsize{8}{9}\selectfont
\setlength{\tabcolsep}{1.0pt}
\renewcommand{\arraystretch}{1.15}
\begin{tabular*}{\textwidth}{@{\extracolsep{\fill}}
                >{\centering\arraybackslash}p{2.75cm}
                c c c c@{\hspace{6pt}}c c c c@{}}
\hline
\multirow{2}{*}{Method} &
\multicolumn{4}{c@{\hspace{6pt}}}{Clinical Inferability} &
\multicolumn{4}{c}{VCD} \\
\cmidrule(r{4pt}){2-5}\cmidrule(l{4pt}){6-9}
& \raisebox{0.6ex}{$S \uparrow$}
& \raisebox{0.6ex}{$C \uparrow$}
& \raisebox{0.6ex}{CII $\uparrow$}
& \raisebox{0.6ex}{$G$}
& \raisebox{0.6ex}{MI $\uparrow$}
& \shortstack{Diabetes\\$\uparrow$}
& \shortstack{Colorectal\\$\uparrow$}
& \shortstack{Average\\$\uparrow$} \\
\hline
gpt-5.6-terra & 0.828 & 0.787 & 0.803 & 0.041 & 0.210 & 0.215 & 0.369 & 0.265 \\
gemini-3.5-flash & 0.822 & 0.789 & 0.796 & 0.032 & 0.250 & 0.390 & 0.265 & 0.302 \\
OpenAI Agents SDK (gpt-5.6-terra) & 0.752 & \textbf{0.821} & 0.769 & $-0.069$ & 0.262 & 0.283 & 0.370 & 0.305 \\
Google ADK (gemini-3.5-flash) & 0.805 & 0.780 & 0.783 & 0.025 & 0.300 & 0.432 & 0.324 & 0.352 \\
\textbf{CaseWeaver} & \textbf{0.930} & 0.803 & \textbf{0.860} & 0.127 & \textbf{0.547} & \textbf{0.620} & \textbf{0.686} & \textbf{0.618} \\
\hline
\end{tabular*}
\endgroup
\end{table}

Figure~\ref{fig:clinical-inferability}(b) further shows the relationship between diagnostic success and information coverage. CaseWeaver maintained high $S$ and $C$, indicating that its evidence was both accessible during interaction and diagnostically informative. In contrast, the OpenAI Agents SDK baseline achieved the highest information coverage ($C=0.821$) but a diagnostic success rate of only 0.752. This high-coverage, low-success pattern suggests that accessible evidence may have lacked sufficient organization or diagnostic direction to support the correct diagnosis. CaseWeaver's multimodal evidence instead supported an interactive reasoning process from information acquisition to diagnosis, indicating stronger clinical inferability.

\subsection{VCD Results}

As a complement to the clinical inferability evaluation, we further assessed VCD. Each method generated 50 virtual cases, comprising 20 myocardial infarction, 20 diabetes, and 10 colorectal cancer cases. CaseWeaver achieved the highest VCD across all three disease tasks, with scores of 0.547, 0.620, and 0.686 and a macro-average of 0.618, compared with 0.265--0.352 for the four baselines (Table~\ref{tab:clinical-vcd-results}). These results indicate greater within-disease variation in patient backgrounds, phenotypes, clinical courses, and evidence plans, reducing concentration around a small number of fixed templates.

\section{Conclusion}

CaseWeaver uses a timeline-anchored LCCG and protocol-driven modality agents to generate interrelated multimodal evidence for the same patient. The AgentClinic and VCD evaluations indicate that the generated cases encompass diverse patient profiles and provide evidence that can be progressively acquired and integrated to support diagnostic reasoning. By linking longitudinal disease progression with multimodal examination findings, CaseWeaver generates patient-level cases with an inferable clinical structure rather than isolated data outputs. The framework may therefore serve as a virtual case generator for medical AI training and clinical education.

\bibliographystyle{splncs04}
\bibliography{references}

\end{document}